\newcounter{myctr}
\def\myitem{\refstepcounter{myctr}\bibfont\noindent\ifnum\themyctr>9\else\phantom{0}\fi\hangindent17pt\themyctr.\enskip}

\newcommand{\ket}[1]{\left\vert#1\right\rangle}

\newcommand{\bra}[1]{\left\langle#1\right\vert}

\documentclass{ws-ijqi}

\usepackage{epstopdf}

\begin{document}

\markboth{Steve Campbell, Laura Mazzola, and Mauro Paternostro}
{Global quantum correlations in the Ising spin model}

\catchline{}{}{}{}{}

\title{GLOBAL QUANTUM CORRELATIONS IN THE ISING MODEL}

\author{STEVE CAMPBELL}
\address{Physics Department, University College Cork, Cork, Ireland \&\\
Centre for Theoretical Atomic, Molecular and Optical Physics, School of Mathematics and Physics, Queen's University, Belfast BT7 1NN, United Kingdom,\\
\ scampbell@phys.ucc.ie}

\author{LAURA MAZZOLA}
\address{Turku Centre for Quantum Physics, Department of Physics and Astronomy, University of Turku, FI-20014 Turun yliopisto, Finland \&\\
Centre for Theoretical Atomic, Molecular and Optical Physics, School of Mathematics and Physics, Queen's University, Belfast BT7 1NN, United Kingdom,\\
\ l.mazzola@qub.ac.uk}

\author{MAURO PATERNOSTRO}
\address{Centre for Theoretical Atomic, Molecular and Optical Physics, School of Mathematics and Physics, Queen's University, Belfast BT7 1NN, United Kingdom}
\maketitle

\begin{history}
\received{Day Month Year}
\revised{Day Month Year}
\end{history}

\begin{abstract}
We study quantum correlations in an isotropic Ising ring under the effects of a transverse magnetic field. After characterizing the behavior of two-spin quantum correlations, we extend our analysis to global properties of the ring, using a figure of merit for quantum correlations that shows enough sensitivity to reveal the drastic changes in the properties of the system at criticality. This opens up the possibility to relate statistical properties of quantum many-body systems to suitably tailored measures of quantum correlations that capture features going far beyond standard quantum entanglement.
\end{abstract}

\section{Introduction}
Quantum many-body systems of interacting spin particles embody a splendid scenario for the study of quantum statistics, the investigation of fundamental questions at the genuine multipartite level and the simulation of non-trivial interaction models of generally difficult {\it natural} accessibility. Moreover, a very successful decade of theoretical efforts has crowned quantum spin systems as promising devices for the realization of computation and short-haul communication protocols  in on-chip solid-state quantum devices\cite{BosePRL03,BoseCont07,DiFrancoPRL08,DiFrancoPRL09}.

Much of the handiness and interest in analyzing chains and rings of interacting quantum spins comes from an extensive body of work performed in the area of so-called {\it exactly solvable models}, where analytical and numerical techniques for the investigation of statistical properties of a vast class of many-body systems have been formalized and put in place\cite{AmicoRMP08,Sachdev,EisertRMP10,GabrieleJCTN}. A nice link between such investigations and a genuine information theoretical viewpoint comes from the study of figures of merit such as entanglement in relation to critical properties of interacting many-body systems dragged across a quantum phase transition (QPT)\cite{AmicoRMP08,OsbornePRA02,OsterlohNat02,GuPRL04,ZanardiPRL06,CormickPRA08,ApollaroNJP10,RossiniPRA07,RossiniPRA08}.
The increasing interest in the understanding of general quantum correlations\cite{OllivierPRL01,HendersonJPA01,OppenheimPRL02,GroismanPRA05,LuoMID,ModiPRL10} and their distinctive features that often depart from those characterizing the latter motivate a research aiming at establishing a connection between statistical properties of quantum many-body systems and general quantum correlations. Some interesting steps have been performed along these lines\cite{WerlangPRL10,MazieroArX10,RossiniArX10}, while the topic still deserves a more systematic development.

We thus present a detailed analysis of the quantum correlations in an Ising spin ring in transverse magnetic field \cite{BuzekPRA04,GunlyckePRA01,MontakhabPRA10} and focus our attention on the global properties of the system. In fact, we aim at establishing a formal parallel between a recent study by some of us on the genuine non-locality content of the ground and thermal state of such a system\cite{CampbellPRA10} and the degree of global quantum correlations set among the spins by the interaction here at hand. We consider different quantifiers of general quantum correlations, each capturing the various aspects of such a multifaceted problem. We go beyond the study of mere two-spin quantum correlations to consider, on the contrary, their multipartite version. To this task, we employ a generalization of quantum discord based on a natural extension of mutual information to the many-spin scenario\cite{RulliArX11}. We demonstrate the behavior of such correlations as a function of the transverse magnetic field when changing the number of spins in the ring. Very interestingly, we show that, while tests for multipartite non-locality fail to reveal the modifications occurring within the system at criticality\cite{CampbellPRA10}, global quantum correlations quite efficiently witness the occurrence of a QPT. While a detailed study of such a relation is left for further work\cite{post}, we believe that our analysis sets the ground for a fertile investigation.

The remainder of the manuscript is organized as follows: in Sec.~\ref{model} we briefly review the Ising model of spin one-half particles arranged in a ring configuration and under the effects of a transverse magnetic field and briefly discuss the technique we use for its diagonalization. In Sec.~\ref{corre} we recall the definition of bipartite and multipartite quantum correlations we are going to use in this paper. Sections~\ref{study} and \ref{studyglobal} are devoted to the study of the behavior of two-spins and global correlations, respectively, in the ground state of the quantum spin chain. Finally, in Sec.~\ref{conc} we draw our conclusions.

\section{The Model}
\label{model}
The model under investigation is an isotropic Ising chain in a transverse magnetic field. For $N$ coupled spins, the nearest-neighbor Hamiltonian model reads
\begin{equation}
\label{model0}
\hat{H}=-J\sum^{N}_{n=1} \hat\sigma^{x}_{n} \otimes \hat\sigma^{x}_{n+1} + B \sum^{N}_{n=1} \hat\sigma^{z}_{n}
\end{equation}
where $J$ ($B$) is a dimensionless parameters representing the inter-spin coupling strength (the global magnetic field) and $\{\hat \sigma^x_n,\hat \sigma^y_n,\hat \sigma^z_n\}$ is the set of Pauli spin operators for spin $n{=}1,..,N$. We assume cyclic boundary conditions such that $\hat\sigma^i_{N+1}=\hat\sigma^i_1~(i{=}x,y,z)$. The model is exactly solvable\cite{OsbornePRA02,BuzekPRA04,MathisAoP61} and here we briefly recall the methodology used to tackle it. In order to diagonalize the Hamiltonian we introduce the raising/lowering operator $\hat \sigma^{\pm}=(\hat \sigma^x_n \pm i\hat \sigma^y_n)/2$, and then move into a new fermionic picture defined by  the Jordan-Wigner transformation\cite{Jordan-Wigner}
\begin{equation}
\hat c^\dag_n=(\hat c_n)^\dag=\otimes^{n-1}_{j=1}(-\hat{\sigma}^z_j)\hat{\sigma}^+_n,~~~~\{\hat c_n,\hat c^\dag_m\}=\delta_{nm},
\end{equation}
where $\delta_{nm}$ is the Kronecker delta. These fermionic variables allow the Hamiltonian to be written
\begin{equation}
\hat H{=}-J\sum^{N}_{n=1}(\hat c^{\dagger}_n-\hat c_n)(\hat c^{\dagger}_{n+1}+\hat c_{n+1})+2B \sum^{N}_{n=1}(\hat c_n^\dagger \hat c_n -\frac{1}{2}).
\end{equation}
A transformation to the momentum representation is then performed using a Fourier transform while full diagonalization is achieved by  final Bogoliubov transformation introducing new fermionic operators $\{\hat b_k,\hat b^\dag_k\}$ ($k=-N/2,..,N/2-1$) and giving a free-fermion Hamiltonian
\begin{equation}
\label{free}
\hat{H}_{\text{ff}}=\sum_{k}\epsilon_k\hat{b}^\dag_k\hat{b}_k- \sum_k\epsilon_k,
\end{equation}
where ${\epsilon_k=\sqrt{J^2+B^2-2JB\cos \phi_k}}$ and ${\phi_k=\pi(2 k+1)/N}$  in the subsector with an even number of fermions and a slightly different definition for the odd-number case\cite{CampbellPRA10}. The ground state of the system is then the state satisfying the eigenvalue equation $\hat b_k\ket{\mathrm{GS}_N}=0~\forall{k}$, which gives us
\begin{equation}\label{GS}
\ket{\mathrm{GS}_N}=\bigotimes_{k}\left(\cos\frac{\vartheta_k}{2}\ket{00}_{k,-k}+i\sin\frac{\vartheta_k}{2}\ket{11}_{k,-k}\right)
\end{equation}
with $\tan\vartheta_k=(-B+J\cos\phi_k)/(J\sin\phi_k)$ and $\ket{0}_{\phi_k}$ ($\ket{1}_{\phi_k}$) the state with no (one) fermion with momentum $\phi_k$. The energy of the ground state is $\Lambda_N=-\sum_k \epsilon_k$.


\section{Bipartite and Multipartite Quantum Correlations}
\label{corre}
In this Section we introduce the main tools used in our investigation. We first review two figures of merit for quantum correlations in bipartite systems: the quantum discord and the (ameliorated) measurement-induced disturbance. We then move to the multipartite scenario and introduce a measure of multipartite non-classical correlations proposed by Rulli and Sarandy in Ref. \cite{RulliArX11}.

\subsection{Quantum discord, measurement-induced disturbance and its improved version}
Quantum discord is based on the idea of quantifying the discrepancy between the generalization to the quantum domain of two expressions for the mutual information\cite{OllivierPRL01}. Imagine to have a bipartite system described by the density operator $\rho_{AB}$ with $\rho_{A}$ ($\rho_{B}$) denoting the reduced state of system A (B). The total correlations between A and B are quantified by the quantum generalization of mutual information
\begin{equation}
I(\rho_{AB})=S(\rho_{A})-S(\rho_{A}|\rho_{B})
\end{equation}
where $S(\rho_{A})=-\mathrm{Tr}\rho_{A}\log_2\rho_{A}$ is the von Neumann entropy and $S(\rho_{A}|\rho_{B})=S(\rho_{AB})-S(\rho_B)$ is the conditional entropy. Nevertheless, by using a measurement-based approach, a second definition of conditional entropy is possible. The application of local projective measurements on a part of the system projects the total system in a different state. In particular if the measurement is described by the set of projectors $\{\Pi_B^j\}$, the conditional density operator (i.e. the state of the total system AB conditioned on the measurement outcome labeled by $j$) is written as $\rho_{AB|j}=(\mathbf{1}_A\otimes\Pi_B^j)\rho_{AB}(\mathbf{1}_A\otimes\Pi_B^j)/p_j$, where $p_j=\mathrm{Tr}[(\mathbf{1}_A\otimes\Pi_B^j)\rho_{AB}]$ is the probability of outcome $j$ and $\hat{\bf 1}$ is the identity operator. One thus define the measurement-based conditional entropy $S(\rho_{AB}|\Pi_B^j)=\sum_j p_j S(\rho_{A|j})$ with $\rho_{A|j}=\text{Tr}[\hat\Pi^j_B\rho_{AB}]/p_j$ and finds the alternative version of the quantum mutual information
\begin{equation}
J(\rho_{AB})=S(\rho_A)-S(\rho_{AB}|\Pi_B^j),
\end{equation}
which is often referred to as one-way classical correlation\cite{HendersonJPA01}. The difference between quantum mutual information and classical correlations, minimized over the whole set of orthogonal projective measurements on B, defines quantum discord as
\begin{equation}
\label{discord}
{\cal D}^{B\rightarrow A}(\rho_{AB})=\inf_{\{\Pi_B^j\}}[I(\rho_{AB})-J(\rho_{AB})].
\end{equation}
It should be noted that the minimization implied in the definition of ${\cal D}^{B\rightarrow A}(\rho_{AB})$ makes its analytical evaluation very difficult. To date, quasi-closed analytic expressions are in fact known only for quite a restricted class of two-spin states\cite{Alber}.
The intrinsic asymmetry of Eq.~(\ref{discord}) can be lifted by considering the symmetrized form ${\cal D}=\max[{\cal D}^{A\rightarrow B}(\rho_{AB}),{\cal D}^{B\rightarrow A}(\rho_{AB})]$, which is null only on so-called classical-classical states\cite{Piani}, i.e. density matrices that can be written as $\sum_{ij}p_{ij}|i,j\rangle\langle{i,j}|$ with $\{\ket{i}\}$ and $\{\ket{j}\}$ single-spin orthonormal sets and $p_{ij}$ a joint probability distribution for indices $(i,j)$, and is thus a faithful indicator of quantum correlations.

In Ref. \cite{LuoMID} Luo introduced measurement-induced disturbance (MID) as a different quantifier of quantum correlations based on the alterations induced on a quantum mechanical state by a measurement process. Under a bilocal complete projective measurement $\{\Pi_i^A\otimes\Pi_k^B\}$, a classical state remains invariant, i.e. $\rho_{AB}\equiv\Pi(\rho_{AB})=\sum_{ik}(\Pi_i^A\otimes\Pi_k^B)\rho_{AB}(\Pi_i^A\otimes\Pi_k^B)$.
On the other hand, any complete local projective measurement and in particular one built on the eigenprojectors $\{\Pi_{E,i}^{A},\Pi_{E,k}^{B}\}$ of the reduced states of A and B, fully decoheres a quantum mechanical state, rendering it just a classical statistical distribution of probabilities. The idea behind MID is thus to quantify non-classicality by evaluating the difference between $\rho_{AB}$ and $\Pi_E(\rho_{AB})=\sum_{ik}(\Pi_{E,i}^A\otimes\Pi_{E,k}^B)\rho_{AB}(\Pi_{E,i}^A\otimes\Pi_{E,k}^B)$. Quantitatively, MID is defined as
\begin{equation}
\mathcal{M}(\rho_{AB})=I(\rho_{AB})-I(\Pi_E(\rho_{AB})).
\end{equation}
Evidently, MID is much easier to compute than quantum discord due to the lack of any optimization procedure over the set of projective measurements, and represents an upper bound to ${\cal D}$. However, such a lack of optimization can lead to inconsistencies between the two indicators: MID can be non-null and even maximal on states exhibiting zero quantum discord. To remove this inconsistency an improved version of MID (AMID) has been proposed\cite{WPM,GirolamiArX10} that includes the {\it ab initio} optimization over any possible set of local projectors on part A and B. Therefore the definition of AMID is
\begin{equation}\label{AMID}
\mathcal{A}(\rho_{AB})=\inf_{\Pi}[I(\rho_{AB})-I(\Pi(\rho_{AB}))]
\end{equation}
with $\Pi=\{\Pi_{i}^{A}\otimes\Pi_{k}^{B}\}$ as before. The quantitative relation between discord and AMID, which is faithful by construction, has been explored  and experimentally demonstrated using a linear optics setup generating a hyperentangled state of four photons\cite{Chiuri2011}.

\subsection{Global quantum discord}
Here we briefly discuss a measure for the global content of non-classical correlations in the state of a multipartite system. By noting that the original definition of discord\cite{OllivierPRL01} can be rewritten in terms of relative entropy\cite{RulliArX11}, the following symmetric extension of discord can be considered
\begin{equation}
\mathcal{D}^{B\rightarrow A}(\rho_{AB})=S(\rho_{AB}||\Pi(\rho_{AB}))-S(\rho_{A}||\Pi_A(\rho_{A}))-S(\rho_{B}||\Pi_B(\rho_{B}))
\end{equation}
where $S(\rho_1||\rho_2)=\mathrm{Tr}[\rho_1\log_2\rho_1-\rho_1\log_2\rho_2]$ is the relative entropy between states $\rho_1$ and $\rho_2$. The global quantum discord $\mathcal{GD}(\rho_{A_{1}...A_{N}})$, which quantifies multipartite non-classical correlations in a system built out of the set of parties $\{A_{n}\}$, is defined as
\begin{equation}\label{GQD}
\mathcal{GD}(\rho_{A_1...A_N})=\inf_{\{\hat\Pi_j\}}\left[S(\rho_{A_1...A_N}||\hat\Pi(\rho_{A_{1}...A_{N}}))-\sum_{j=1}^N S(\rho_{A_j}||\hat\Pi_j(\rho_{A_j}))\right],
\end{equation}
where $\hat\Pi_j(\rho_{A_j})=\sum_{j'}\hat\Pi_{A_j}^{j'}\rho_{A_j}\hat\Pi_{A_j}^{j'}$ and $\hat\Pi(\rho_{A_{1}...A_{N}})=\sum_k \hat\Pi_k \rho_{A_{1}...A_{N}} \hat\Pi_k$ with $\hat \Pi_k=\otimes^N_{l=1}\hat \Pi^{j_l}_{A_l}$ and $k$ denoting the index string $(j_1... j_N)$. Eq.~(\ref{GQD}), where the infimum is taken over all possible multi-local projectors $\hat \Pi_j$, is always non-negative but its maximum value depends on the dimension of the total Hilbert space at hand.

\section{Two-spin Quantum Correlations in the Ground State of an Ising Ring}
\label{study}
In this Section we study the behavior of the two-spin quantum correlations. We fix the strength coupling parameter $J$ and investigate the quantum correlations shared by pairs of spins ``extracted" from the ring against the magnetic field $B$. We also study the behavior of quantum correlations when the ring increases in size and pairs made out of non-nearest-neighbor spins are considered. We compare the indications provided by the various figures of merit for two-qubit non-classical correlations discussed in the previous Section and highlight a series of interesting features.

As here we are focusing our attention to two-spin quantum correlations, we can take advantage of the fact that the density matrix of any pair of spins $(i,j)$ can be expressed by means of two-point correlation functions as\cite{OsbornePRA02}
\begin{equation}
\rho_{ij}=(\hat{\bf 1}_{4}+\!\!\!\!\sum_{a,b=0,x,y,z}\!\!\!\!{\chi}^{ab}_{ij}\hat\sigma^a_i\otimes\hat{\sigma}^b_j)/4
\end{equation}
with ${\bm \chi}$ the two-point correlation matrix with entries ${\chi}^{ab}_{ij}=\langle\hat\sigma^a_i\otimes\hat{\sigma}^b_j\rangle$ and $\hat\sigma_i^{0}{\equiv}\hat{\bf 1}$. In general, getting the expressions of such correlators is a difficult task due to the non-local form of the ground state in the Jordan-Wigner representation\cite{OsbornePRA02}. For completeness of presentation, we should however mention that in the thermodynamic limit of $N\rightarrow\infty$, two-point correlations can be expressed in terms of determinants of Toeplitz matrices. The symmetries enjoyed by $\hat H$ are such that the only non-zero elements of ${\bm\chi}$ are
\begin{equation}
\begin{aligned}
&\chi^{xx}_{ii+s}
\left|
\begin{matrix}
G_{-1}&G_{-2}&\cdots&G_{-s}\\
G_{0}&G_{-1}&\cdots&G_{-s+1}\\
\vdots&\vdots&\ddots&\vdots\\
G_{s-2}&G_{s-3}&\cdots&G_{-1}
\end{matrix}
\right|,
~~~
\chi^{yy}_{ii+s}
\left|
\begin{matrix}
G_{1}&G_{0}&\cdots&G_{-s+2}\\
G_{2}&G_{1}&\cdots&G_{-s+3}\\
\vdots&\vdots&\ddots&\vdots\\
G_{s}&G_{s-1}&\cdots&G_{1}
\end{matrix}
\right|,\\
&\chi^{zz}_{ii+s}=\frac{1}{\pi^2}\left(\int^\pi_0d\phi\frac{1+\lambda\cos\phi}{\tilde\epsilon}\right)^2-G_sG_{-s},
\end{aligned}
\end{equation}
where $s$ is an integer representing the number of sites of the ring separating the two elements of the pair being considered. By recasting the Hamiltonian model as $\hat H=\sum^N_{n=1}(-\lambda\hat\sigma^x_n\otimes\hat\sigma^x_{n+1}+\hat\sigma^z_n)$, we have\cite{Baruch}
\begin{equation}
G_k=\frac1\pi\int^\pi_0d\phi\cos(k\phi)\frac{1+\lambda\cos\phi}{\tilde\epsilon}-\frac{\lambda}{\pi}\int^\pi_0d\phi\sin(k\phi)\frac{\sin\phi}{\tilde\epsilon}
\end{equation}
with $\lambda=J/B$ and $\tilde\epsilon=\sqrt{1+\lambda^2+2\lambda\cos\phi}$. As long as two-spin states are taken into account, the
analysis of quantum correlations can be done by directly working on the two-spin reduced density matrix $\rho_{ij}$ and using the equations above, for arbitrary size of the system. In this work, though, we are interested in global properties of the system's state, which in principle require arbitrary multipoint correlation functions, not achievable through the apparatus described above. while the considerations to be brought forward in this Section will be easily generalized to arbitrarily sized systems, we will consider only finite-length systems to illustrate the key points of our study and set useful benchmarks for the analysis presented in Sec.~\ref{studyglobal}. We start our analysis by comparing the results coming from the use of the three quantifiers introduced above. In particular we consider the quantum correlations contained in the reduced state of two neighboring spins as measured by quantum discord, MID and AMID. In what follows, without affecting the generality of the analysis, we set $J=1$ in the model in Eq.~(\ref{model0}) and leave $B$ as a free coefficient. In fact, the relevant parameter in the dynamics under scrutiny is the ratio $B/J$.

A remark is due in respect to the computation of some of the figures of merit addressed here. It turns out that the reduced density matrix of any pair of spins in our problem can be written, regardless of the value of the magnetic field and the coupling strength, in the {\it X-like form}
\begin{equation}
\rho_{ij}=
\begin{pmatrix}
\star&&&\star\\
&\star&\star&&\\
&\star&\star&&\\
\star&&&\star\\
\end{pmatrix}
\end{equation}
with $\star$ indicating the only non-zero elements of the density matrix.  
In this case, differently from what occurs for general two-qubit states, a semi-closed analytic formula for the evaluation of ${\cal D}^{j\rightarrow i}(\rho_{ij})$ is available\cite{Alber} (notice that ${\cal D}^{j\leftarrow i}(\rho_{ij})$ can be easily calculated with the very same formula by first applying a Swap gate to $\rho_{ij}$). While we point the reader to Ref.\cite{Alber} for full details on this, here it is enough to mention that we have used such formula for the calculations reported in our work, thoroughly checking the corresponding predictions with an exact numerical approach.  As for MID, the lack of optimization over the local projective bases makes its evaluation straightforward and no further comment is needed in this respect. Finally, AMID is calculated exploiting once more the X-like form of the reduced two-spin states and the formula found for this task in Ref.\cite{GirolamiArX10} (as for discord, we have duly checked the consistency of the analytic predictions with those of a fully numerical study).

\begin{figure}[t]
\centerline{\includegraphics[width=9cm]{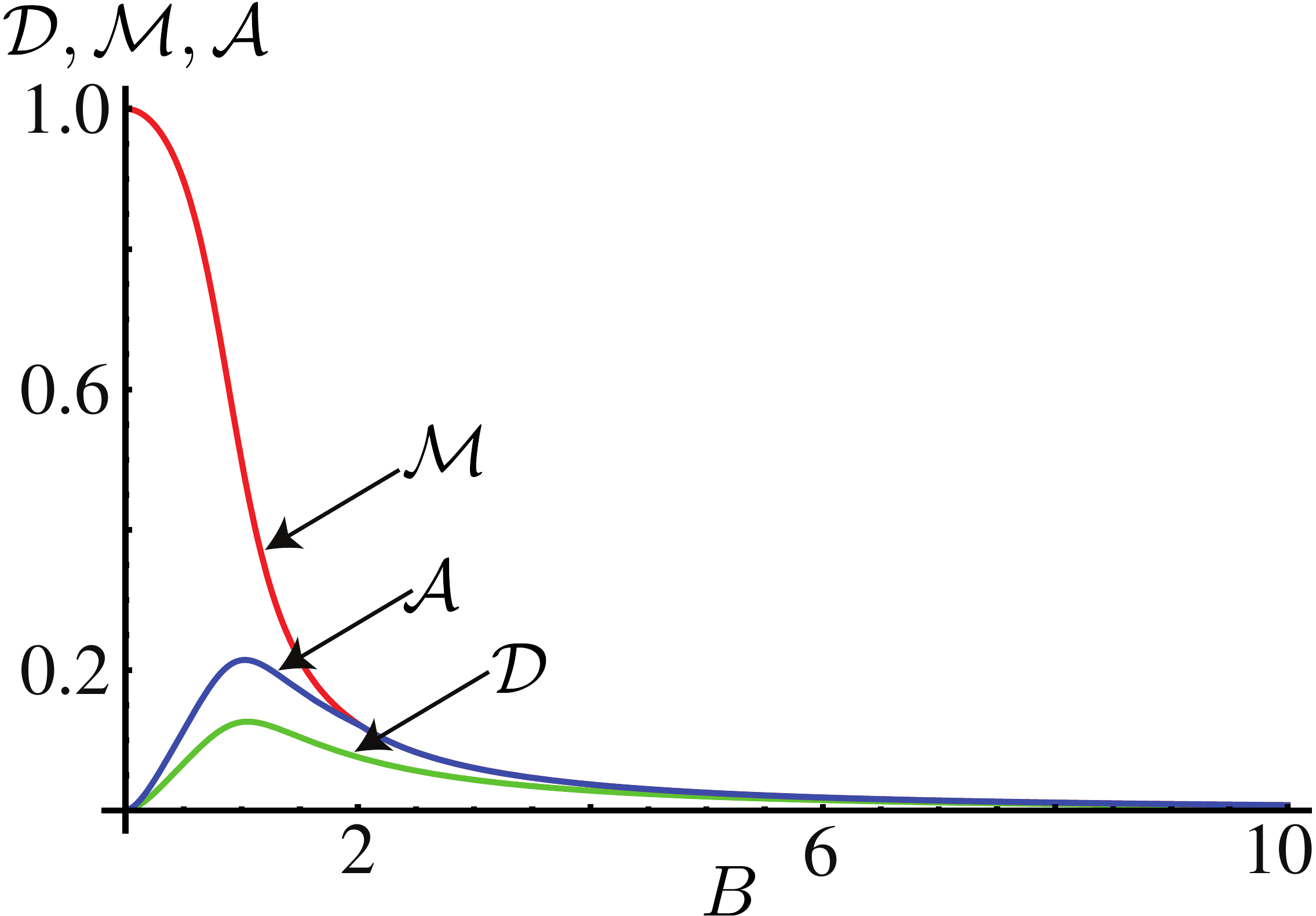}}
\caption{Quantum correlations shared by nearest neighbors in a chain of $N=6$ spins plotted against $B$ for $J=1$. The plot shows the clear inconsistency between MID and more faithful measures of non-classical correlations such as ${\cal D}$ and ${\cal A}$ and the upper-bound to discord embodied by AMID.}\label{MID}
\end{figure}

 In Fig.~\ref{MID} we present the results corresponding to the case of a ring of $N=6$ spins.
At $B=0$, the reduced state of any two spins $\alpha,\beta$ (with $\alpha,\beta=1,2,..,N$), obtained as $\rho_{\alpha,\beta}=\text{Tr}_{\{N'\}}(\ket{\text{GS}_N}\bra{\text{GS}_N})$ (with $\{N'\}=\{1,2,..,N\}\backslash \{\alpha,\beta\}$) is an equally-weighted mixture of $\ket{\phi_+}=(\ket{00}+\ket{11})/\sqrt2$ and $\ket{\psi_+}=(\ket{10}+\ket{01})/\sqrt2$. While both discord and AMID predict no quantum correlation in such a class of states, ${\cal M}=1$. One can ascribe such a striking inconsistency of results to the fact that, for such a class of states, the reduced single-spin states are proportional to the identity operator $\hat{\bf 1}$. This implies that the eigenprojections are undetermined, exposing the evaluation of MID to a rather rough overestimation. As soon as $B{>}0$, the balance between such two state components is lost and a proper basis of reduced eigenprojectors can indeed be found. Nevertheless, MID quite considerably tops both ${\cal D}$ and ${\cal A}$. This demonstrates that MID fails to capture the genuine content of quantum correlations even in such a simple two-qubit state and, as such, does not embody a faithful figure of merit for our investigation. On the other hand, ${\cal A}$ provides a much more reliable test, being strictly faithful on classical-classical states and only slightly overestimating the symmetrized discord (it should be stressed that the differences between such measures rise from the fact that they address two different questions, from an operational viewpoint).

As it will be clarified later on, the main point of our analysis is the global content of quantum correlations in the ground state of the Ising ring. Although an extension of AMID to the multipartite scenario is certainly possible, and is quite naturally entailed in the structure of such a figure of merit, the availability of a global quantum discord as a plausible multi-spin indicator and the fact that ${\cal A}$ only constitutes an upper bound to quantum discord push us to consider the latter as our key tool for the remainder of our analysis.

\begin{figure}[t]
\centerline{\includegraphics[width=9cm]{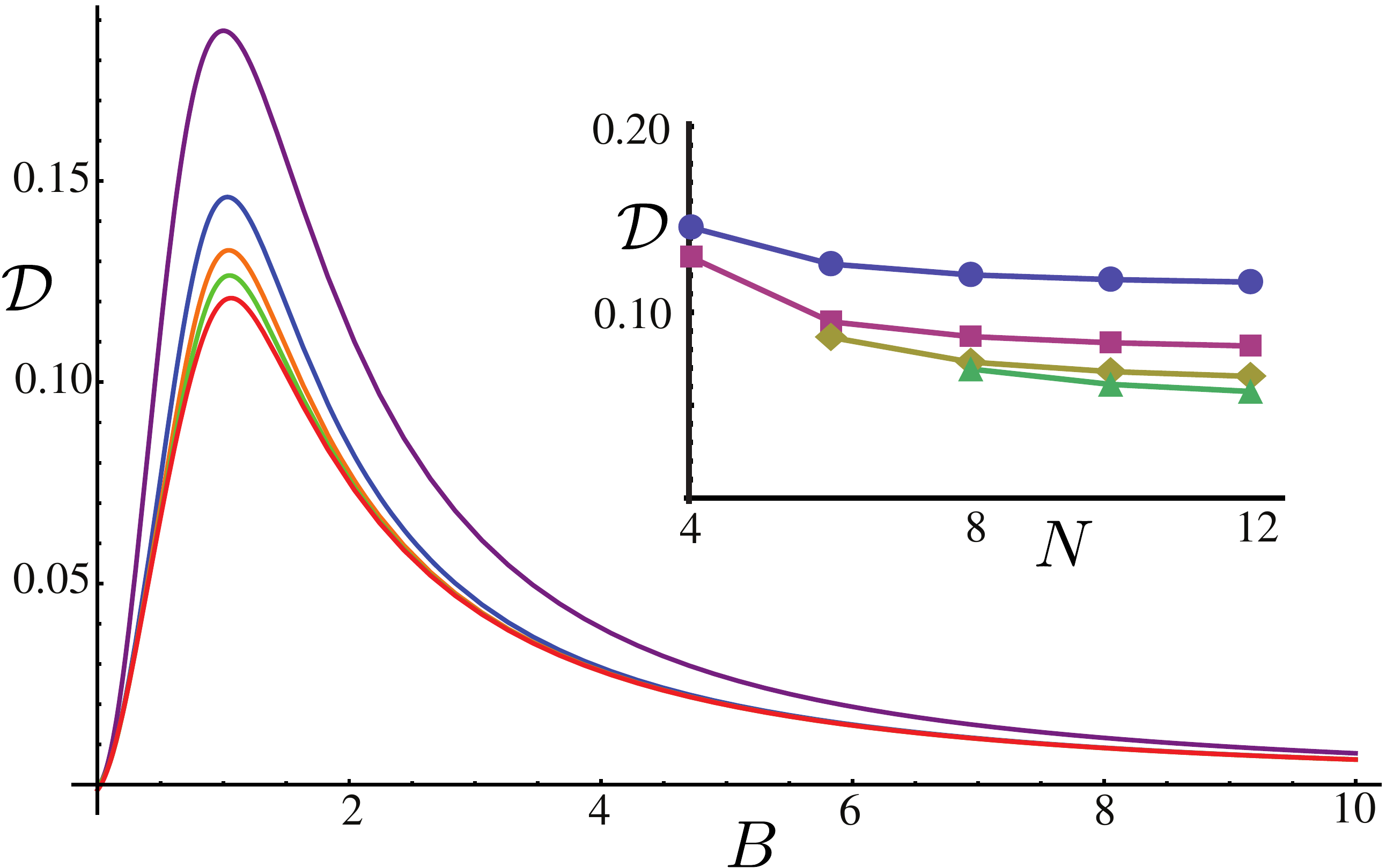}}
\vspace*{8pt}
\caption{(Color online) Quantum discord shared by nearest neighbors in a chain of $N=3, 4, 5, 6$ and 8 spins (from top to bottom curve) against the magnetic field $B$. In the inset we show  quantum discord evaluated at $J/B=1$ against the number of spins in the chain between (i) nearest neighbors (blue line/circled dot); (ii) next-nearest neighbors (purple line/squared dot); (iii) third nearest-neighbors (yellow line/rhombus dot); (iiii) fourth nearest-neighbors (green line/triangular dot)}\label{IncreasingN}
\end{figure}

We now consider how the quantum correlation content of the reduced state of two neighbors depends on the number of spins constituting the whole chain. In Fig.~\ref{IncreasingN} we show how discord in the state of nearest neighbors changes when the number of spins in the ring grows. Evidently, regardless of the number of sites in the spin ring, the region around $B\sim J$ is {\it special}, as the maximum of shared quantum correlations can be found within it. From a statistical mechanics viewpoint, the valence of such a configuration of parameters is quite understandable as it corresponds to the region where the correlation length across the ring diverges. Moreover, using an analysis based on the first derivative of concurrence and discord with respect to $B$\cite{OsbornePRA02,OsterlohNat02,DillenschneiderPRB08}, it has been predicted that criticality emerges at $B=J$. As we will show in the next Section, the special nature of this region extends to global general quantum correlations. 

\begin{figure}[b]
\centerline{\includegraphics[width=9cm]{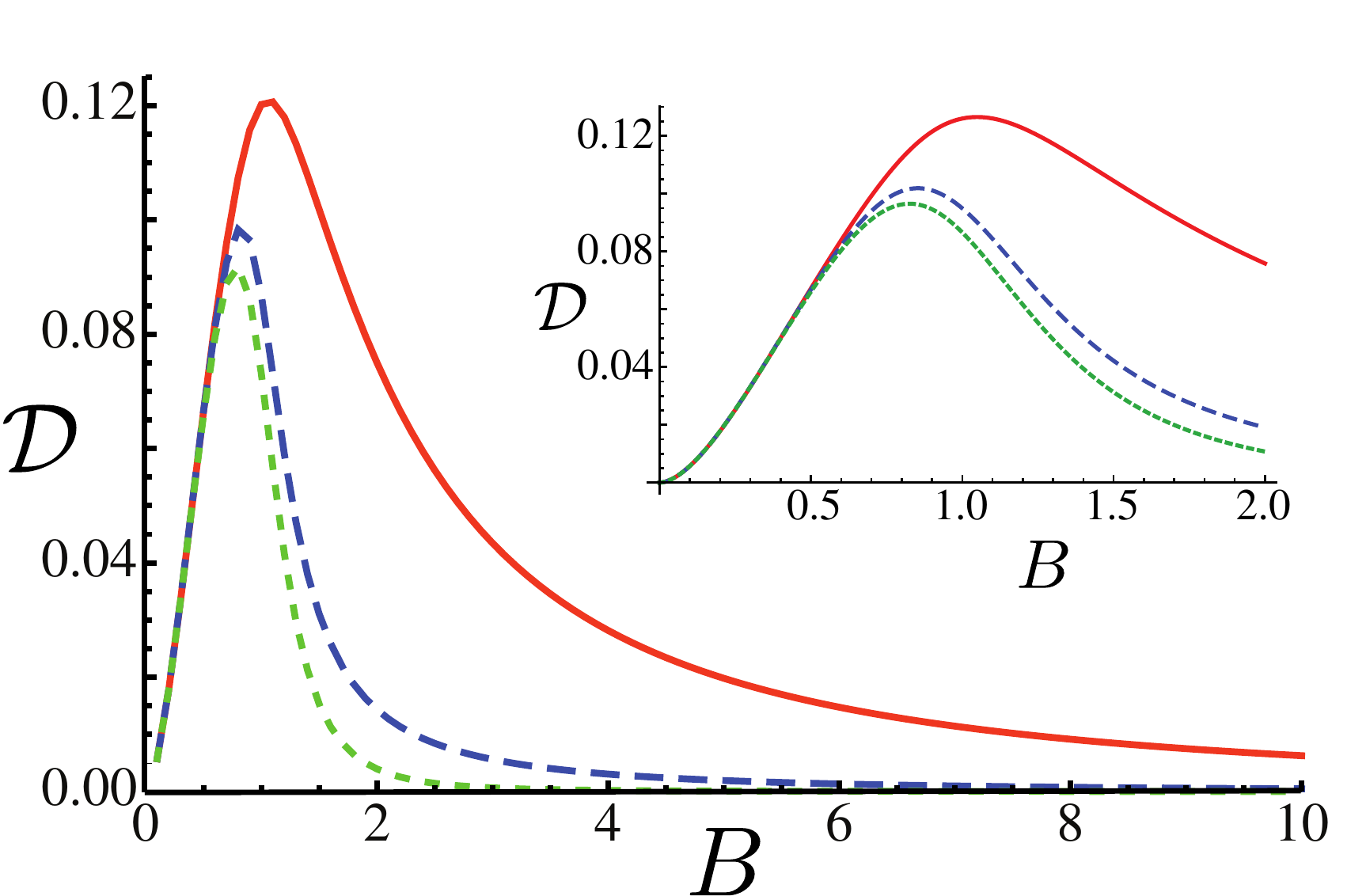}}
\caption{(Color online) Quantum discord as a function of the magnetic field between nearest neighbors (red solid line), next-nearest neighbors (blue dashed line), third nearest-neighbors (green dotted line) in a chain of 8 spins with $J=1$. Inset: Magnification of the main panel in the region of $B\in[0,2]$.}
\label{DiffBip}
\end{figure}

The inset of Fig.~\ref{IncreasingN} shows ${\cal D}$ in increasingly long rings at a set value of the magnetic field ($B=J=1$) and for spins separated by a growing number of sites (from zero up to three). The larger the separation, the smaller the value of discord, although the decrease is rather weak, in stark contrast with what happens to two-spin entanglement in the very same spin model\cite{OsbornePRA02}. Moreover, in Fig.~\ref{DiffBip} we find that the position of the peaks revealed above changes with the site-separation\cite{DillenschneiderPRB08}.

\section{Global Quantum Correlations in the Ground State of an Ising Ring}
\label{studyglobal}
We now tackle the central part of our study, i.e. the analysis of global quantum correlations in the spin model at hand. As anticipated, we make use of the global quantum discord proposed in Ref.\cite{RulliArX11}. When approaching the evaluation of ${\cal GD}$, one faces the problem encompassed by the calculation of the relative entropy between the ground state $\ket{GS_N}$ of the ring and its locally-projected version. For large rings, this could be computationally demanding, let alone the necessity of a global optimization over any possible local projective basis. We thus break the calculation in a few intermediate steps and algebraic rearrangements that help us in streamlining the evaluation of ${\cal GD}$.

First, we consider each single-spin projector $\hat\Pi^j_{n}$ as resulting from the application of a rotation $\hat R_{n}(\theta_j,\varphi_j)$ on the projectors onto the eigenbasis of $\hat\sigma^z_{n}$, i.e. $\{\hat{p}_1=\ket{\uparrow}\bra{\uparrow},\hat{p}_0=\ket{\downarrow}\bra{\downarrow}\}_n$. This leaves us with the observation that the post-measurement state of the ring can be rewritten as
\begin{equation}
\label{prova}
\begin{aligned}
\hat\Pi(\ket{GS_N}\bra{GS_N}){=}\sum^{2^{N}-1}_{k=1}\hat{\cal R}(\{\theta,\varphi\})\hat{\cal P}_k\ket{GS'_N(\{\theta,\varphi\})}\!\bra{GS'_N(\{\theta,\varphi\})}\hat{\cal P}_k\hat{\cal R}^\dag(\{\theta,\varphi\}),\end{aligned}
\end{equation}
where $\ket{GS'_N(\{\theta,\varphi\})}\!\bra{GS'_N(\{\theta,\varphi\})}=\hat{\cal R}(\{\theta,\varphi\})\ket{GS_N}\!\bra{GS_N}\hat{\cal R}^\dag(\{\theta,\varphi\})$ is the rotated ground state of the ring, $\hat{\cal R}(\{\theta,\varphi\})$ is the tensor product of all the rotation matrices needed to produce the local projectors, $\hat{\cal P}_k$ is the tensor product of projectors over the local $\hat\sigma^z$ eigenbasis and we have used the notation $\{\theta,\varphi\}$ as a short-cut to indicate the whole set of angles that enter the rotations. Clearly, the value of label $k$ determines the {\it combination} of $\hat\sigma^z$ eigenprojectors $\{\hat{p}_0,\hat{p}_1\}$ to use in order to build up the $N$-spin projector $\hat{\cal P}_k$. In any case, it is obvious from Eq.~(\ref{prova}) that the application of each $\hat{\cal P}_k$ on the rotated ground state picks up a single diagonal element $\lambda_k$ of the latter so that
\begin{equation}
\hat{\cal P}_k\ket{GS'_N(\{\theta,\varphi\})}\!\bra{GS'_N(\{\theta,\varphi\})}\hat{\cal P}_k{=}\lambda_k\hat{\cal P}_k=\lambda_k\ket{k}\bra{k}
\end{equation}
with $\ket{k}$ the $N$-spin eigenstate of $\otimes^N_{n=1}\hat\sigma^z_n$ determined by our choice of $k$. The set of $\lambda_k$'s, which is straightforward to determine even for a large density matrix, embodies the eigenspectrum of $\hat\Pi(\ket{GS_N}\bra{GS_N})$. The latter is obviously diagonal in the rotated basis $\{\ket{k(\{\theta,\varphi\})}=\hat{\cal R}(\{\theta,\varphi\})\ket{k}\}$. By arranging such vectors in columns, we form the passage matrix $T$ such that $T\hat\Pi(\ket{GS_N}\bra{GS_N})T^\dagger{=}\Lambda{=}\text{diag}[\lambda_0,\lambda_1,..,\lambda_{2^{N}-1}]$, which allows us to write
\begin{equation}
S(\ket{GS_N}\bra{GS_N}{||}\hat\Pi(\ket{GS_N}\bra{GS_N})){=}\!\!\sum^{2^N-1}_{k=1}\mu_k\log_2\mu_k{-}\text{Tr}\left[(T^\dag\ket{GS_N}\bra{GS_N}{T})\log_2\Lambda\right]
\end{equation}
with $\mu_k$ the $k^{\text{th}}$ eigenvalue of $\ket{GS_N}\bra{GS_N}$. Despite its innocence, this expression simplifies the evaluation of the global quantum discord. Needless to say, we still face the necessity for the optimization implied in the definition of ${\cal GD}$. However, the evaluation of its expression prior to this step is now reduced to a computationally non-demanding problem, and this opens up the possibility to explore the thermal-state scenario where the Ising ring is not prepared in its ground state but is affected by the influences of a non-zero temperature\cite{post}.

\begin{figure}[t]
\centerline{{\bf (a)}\hskip4cm{\bf (b)}\hskip4cm{\bf (c)}}
\centerline{\includegraphics[width=12cm]{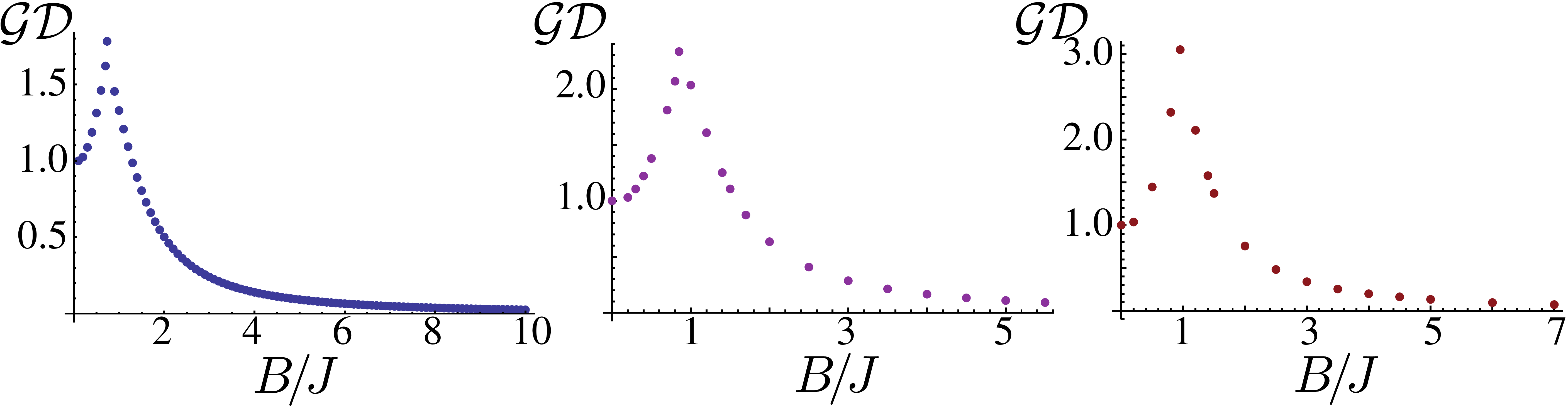}}
\caption{Global quantum discord against the ratio $J/B$ for Ising rings of increasing size. We have taken $N=3$ [panel {\bf (a)}], $4$ [panel {\bf (b)}] and $5$ [panel {\bf (a)}]. The ${\cal GD}$ curve at assigned $N$ is a universal function of $B/J$, peaking in a region around the critical point of the transverse model we are considering. }
\label{main}
\end{figure}

In what follows, we restrict our attention to the ground-state case and investigate the behavior of ${\cal GD}$ against the magnetic term $B$. The results of our calculations for $N=3,4,5$ spins are shown in Fig.~\ref{main} where we have studied ${\cal GD}$ against the ratio $B/J$ so as to investigate universality features of such figure of merit with respect to this parameter.
At small values of $B/J$, the ground state of the ring is locally equivalent to an $N$-spin GHZ state\cite{BuzekPRA04}, for which ${\cal GD}{=}1$\cite{RulliArX11}. By increasing the ratio $B/J$, the system undergoes a profound change in the way correlations (not just entanglement) are shared, as it goes from such a genuinely multipartite entangled state to a fully separable one achieved for $B/J\rightarrow\infty$. The global discord spectacularly captures the special changes in the correlations among the spins occurring close to $J{=}B$ and signals it with a singularity. The functional form of ${\cal GD}$ is independent of the size of the ring, although the decay of collective quantum correlations is faster for rings of smaller size. Due to the handiness of the analytical results described above, we have been able to numerically sample the behavior of ${\cal GD}$ for $N$ up to $8$, finding results in qualitative agreement with the analysis above. The sharp peak in the proximity of $B/J=1$ suggests the existence of a singularity of ${\cal GD}$ at criticality. In turn, this would imply a discontinuous first derivative, thus suggesting a second order phase transition, in agreement with what is known for the Ising model. The trend followed by global discord contrasts quite evidently with what has been found in terms of multipartite non-locality\cite{CampbellPRA10}: the degree of violation of multipartite Bell-like inequalities is a monotonically decreasing function of the magnetic field, thus signaling the profound differences between general quantum correlations and non-local ones.

\begin{figure}[t]
\centerline{{\bf (a)}\hskip6cm{\bf (b)}}
\includegraphics[width=6cm]{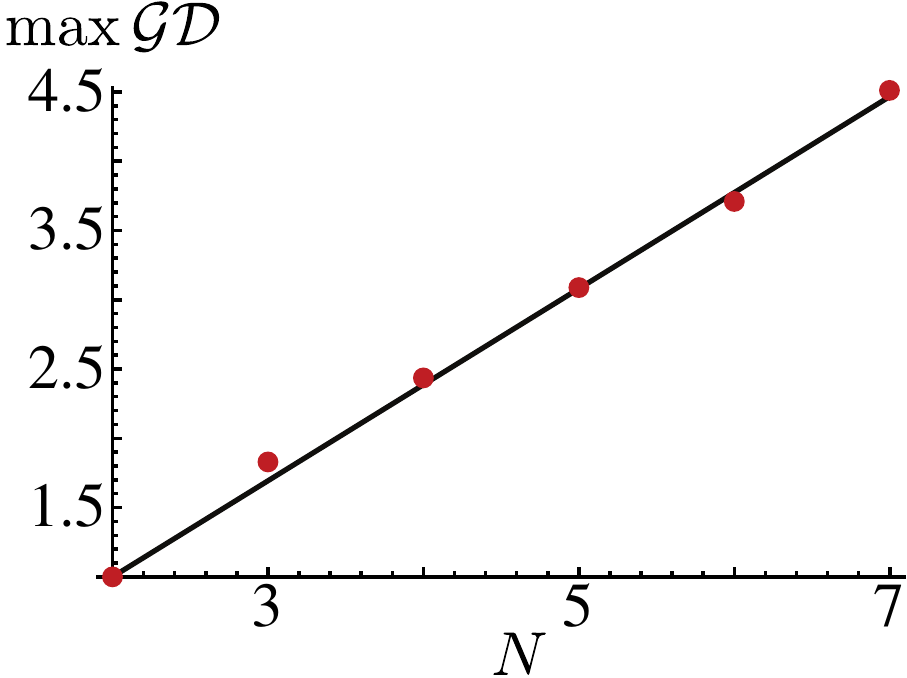}~~~~\includegraphics[width=6cm]{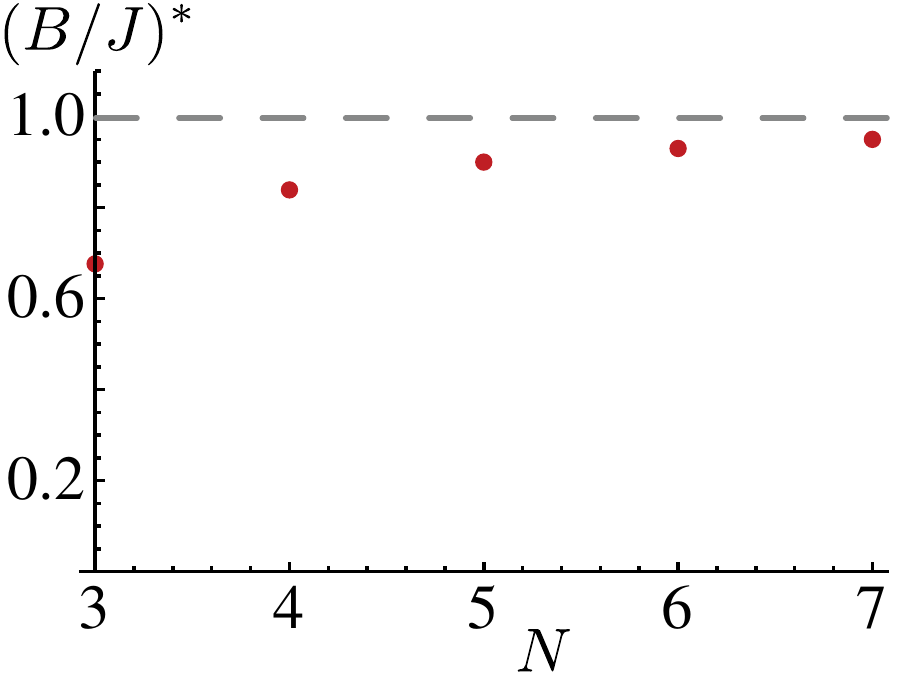}
\caption{(Color online) {\bf (a)} We study the scaling law regulating the maximum value of global discord $\max{\cal GD}$ versus the increasing dimension of the spin ring. The dots show the numerical results of our calculations, while the line shows the best linear fit compatible with the constraint $\max{\cal GD}{\equiv}{\cal D}^\leftrightarrow{=}1$ for $N{=}2$. See the body of the paper for details. {\bf (b)} Value of the magnetic field $(B/J)^*$ at which $\max{\cal GD}$ is achieved, plotted against $N$.}
\label{Scale}
\end{figure}

We have further analyzed the features of ${\cal GD}$ by looking for some qualitative indications on the scaling law regulating the growth of the peak of global discord in the region around ${B/J=1}$. By imposing the constraint $\max{\cal GD}{\equiv}{\cal D}^\leftrightarrow{=}1$ at $N=2$, the numerical results of our calculations appear to be well fitted by the linear function
\begin{equation}
\max{\cal GD}=m(N-2)+1
\end{equation}
with $m$ the slope of the line. We have thus found the best-fit line for the numerical data associated with $N=2,..,7$ provided in Table~\ref{tavolozza}, which gives $m{=}0.693461$ [a comparison between the best-fit function and the numerical data points is given in Fig.~\ref{Scale} {\bf (a)}]. On the other hand, the finite-size nature of the examples worked out here induces some deviations of the value $(B/J)^*$ of the magnetic field at which the singularity of ${\cal GD}$ occurs from the expected criticality point. Such deviations reduce with $N$ growing [cf. Fig.~\ref{Scale} {\bf (b)}] as a power-law behavior.
\begin{table}[b]
\ttbl{30pc}{Dimension $N$ of the spin ring and associated value of $\max{\cal GD}$}
{\begin{tabular}{clccccc}\\
\multicolumn{6}{c}{} \\
\hline
\hline
{$N$}   &{2} &3 &4 &5 &6& 7\\ \hline \hline
$\max{\cal GD}$ &1 &1.8296 &2.4360&3.0879&3.7095&4.501    \\
\hline\hline
\end{tabular}}
\label{tavolozza}
\end{table}

In order to build a parallel with the analysis and the similarities between two-spin entanglement and discord, it is interesting to compare the behavior of such a global measure of quantum correlations with the predictions of a multipartite entanglement measure. To do this, we need a quantum statistical indicator that is able to capture the occurrence of  a quantum phase transition. To this purpose, we adopt the measure for pure states introduced in Ref.\cite{florio}. Given a composite system that we divide in the bipartition $(A,B)$, we define the measure of entanglement
\begin{equation}
\label{eqflorio}
{\cal E}=-\log_2[{\text{Tr}}\rho^2_{A}]
\end{equation}
with $\rho_A={\text{Tr}}_B\ket{\Psi}_{AB}\bra{\Psi}$ the reduced state of subsystem $A$ ($\ket\Psi_{AB}$ is a pure state of the whole system). As $1/{\text{Tr}}\rho^2_{A}$ is viewed as the number of terms entering the Schmidt decomposition of $\ket\Psi_{AB}$, ${\cal E}$ is operatively interpreted as the number of spins effectively entangled (relatively to the considered bipartition)\cite{florio}. For a completely separable state ${\text{Tr}}\rho^2_{A}{=1}$, regardless of the bipartition, so that ${\cal E}{=}0$. The general scenario will see such an indicator depend on the choice of $(A,B)$ and it is intuitive to consider the statistical average $\overline{\cal E}$ of the values achieved by Eq.~\eqref{eqflorio} achieved by exploring all the possible bi-splitting of a system as a measure of the multipartite entanglement shared by its elements. Being this measure statistical in nature, the predictions arising from it would be strongly dependent also on higher statistical moments. The variance of the distribution of entanglement across any possible bi-splitting, in this respect, provides an indication of the sensitivity of the entanglement-sharing to the particular bipartition being taken. A large variance signifies a strong dependence of ${\cal E}$ on the choice of splitting.

\begin{figure}[t]
\centerline{{\bf (a)}\hskip4cm{\bf (b)}\hskip4cm{\bf (c)}}
\centerline{\includegraphics[width=4cm]{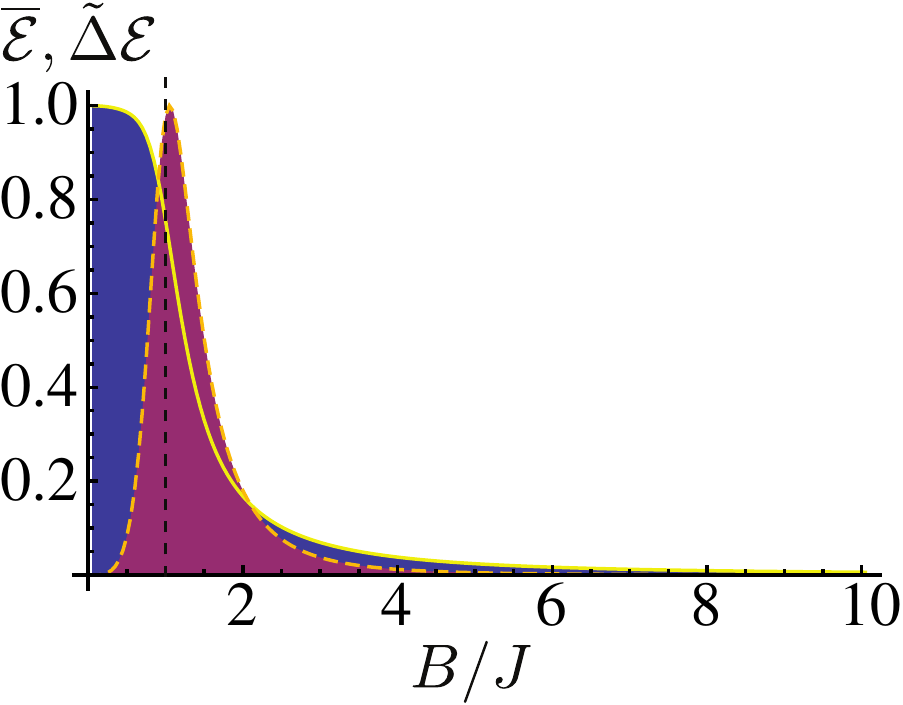}~\includegraphics[width=4cm]{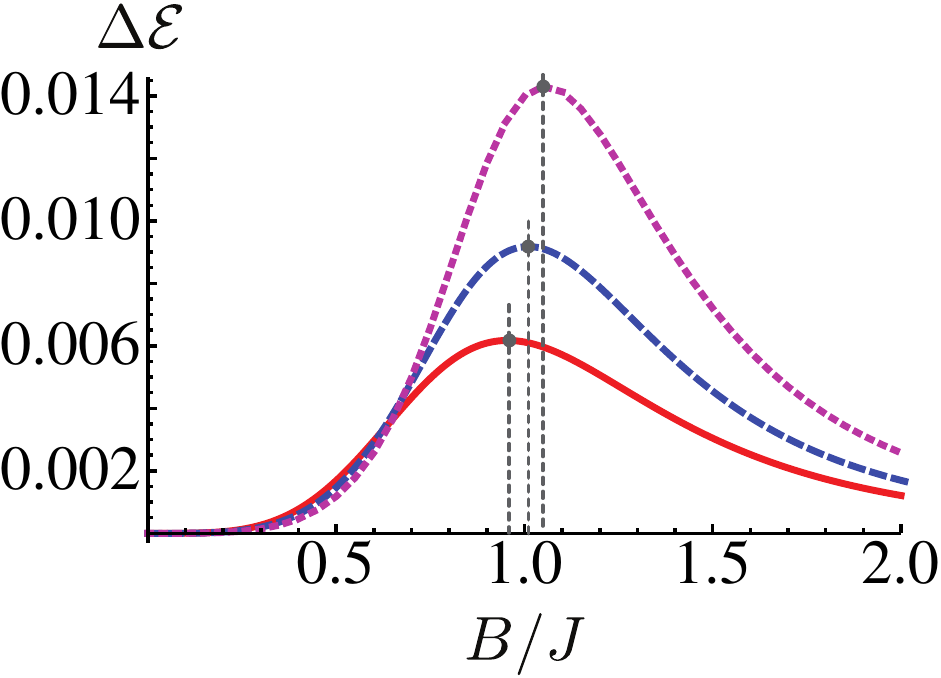}~\includegraphics[width=4cm]{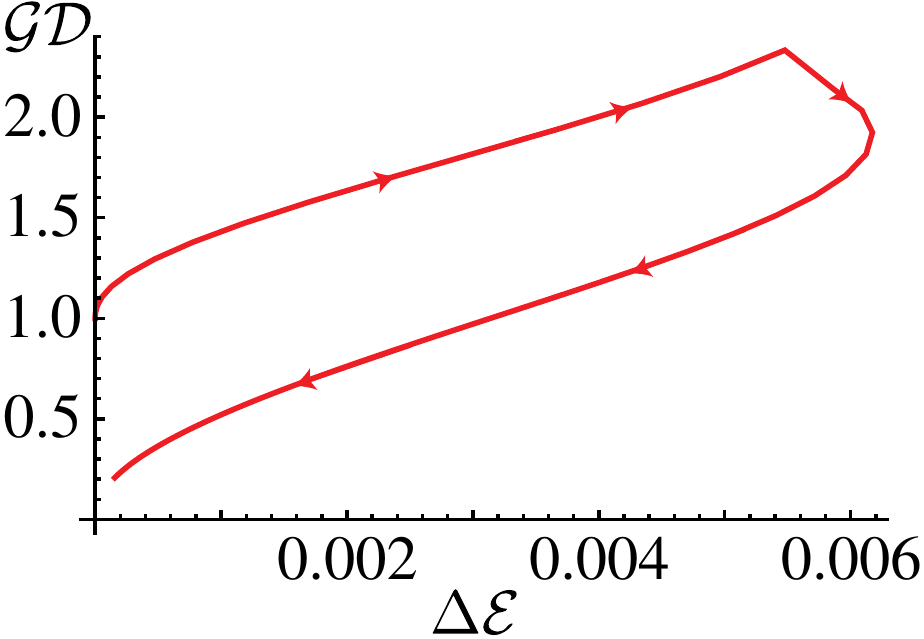}}
\caption{{\bf (a)} Partition-averaged entanglement $\overline{\cal E}$ [solid line] and its normalised variance $\tilde\Delta{\cal E}$ [dashed one] for a ring of six spins plotted against the ratio $B/J$. The vertical line marks the point $B{=}J$. {\bf (b)} Comparison among the variances $\Delta{\cal E}$ (as functions of $B/J$) of the entanglement distributions for a ring of size $N=4,5,6$ (solid, dashed and dotted line, respectively). We mark the position of the maximum of each curve with a dashed vertical line. {\bf (c)} Comparison between ${\cal GD}$ and $\Delta{\cal E}$. As $B/J$ increases, we move along the curve as shown by the arrows.  }
\label{FlorioPlot}
\end{figure}

When the apparatus described above is applied to the state of the spin-ring system, the typical behavior of $\overline{\cal E}$ and the normalized variance $\tilde\Delta{\cal E}=\Delta{\cal E}/(\max\Delta{\cal E})$ is displayed in Fig.~\ref{FlorioPlot} {\bf (a)} (the specific case shown in the figure is relative to $N=6$). For vanishing values of $B/J$, $\overline{\cal E}{=}1$ with a zero-width distribution of entanglement across the possible bipartitions, a situation that clearly witnesses the GHZ nature of the spin-ring state (whose bipartite entanglement is uniformly distributed across the various bi-splitting and this achieves $\Delta{\cal E}{=}0$). Equally expectedly, for large values of $B/J$ no entanglement should be found in the system, regardless of the bipartition. Again, this is well signalled by our analysis, which gives a vanishing $\overline{\cal E}$ with a quickly decaying associated variance, as the system tends to a fully separable state. The intermediate region of values of $B/J$ is the most interesting one: while $\overline{\cal E}$ smoothly decreases without exhibiting any special behavior (except changing from concave to convex),  its variance peaks in the region where $B{\sim}J$ [cfr. the vertical marker located at $J=B$ in Fig.~\ref{FlorioPlot} {\bf (a)}]. Physically, this signifies an increased sensitivity of the entanglement-sharing structure to the specific bi-splitting we consider, thus marking the quick departure from the regular entanglement distribution typical of a GHZ state. Going from an ordered {\it phase} (the GHZ one) to a differently ordered one (corresponding to fully separable states) as $B$ increases, the system is forced to readjust its inner correlations, breaking the symmetry of the way quantum correlations are distributed among its parties and thus increasing the associated variance. This analysis reinforces the idea highlighted above in terms of the behavior of global discord, that such a phase readjustment is a genuinely global phenomenon that is very well captured by an equally global indicator of quantum correlations.

We should notice that the amplitude of the curve describing $\Delta{\cal E}$ is a function of the ring size $N$, and so is the location of $\max\Delta{\cal E}$, as it is illustrated in Fig.~\ref{FlorioPlot} {\bf (b)} for $N=4,5,6$. In analogy with the results displayed in Fig.~\ref{Scale} {\bf (a)}, the maxima are well fitted by a linear function of $N$ and the shifts in their position decreases with $N$. Such similarities call loud for a more direct comparison between ${\cal GD}$ and ${\Delta}{\cal E}$, which is shown in Fig.~\ref{FlorioPlot} {\bf (c)}, where the two figures of merit are plotted against each other (for $B/J\in[10^{-6},6]$, growing as indicated by the sense of the arrows long the curve and, for easiness of calculation, $N=4$). The corresponding open hysteresis path shows very clearly the sharpness of the peak of ${\cal GD}$ and a small (finite-size induced) mismatch between the positions of the maxima of global discord and ${\Delta{\cal E}}$. Such relative shift decreases as $N$ grows, thus signalling an increasing accuracy in determining the actual position of the critical point as the size of the ring increases.

On one hand, the study above confirms the predictions coming from the use of global discord. On the other hand, it is remarkable in showing that differently from quantum discord, which is very efficient in pin-pointing the structural phase transition of quantum correlations in the system, the actual {\it quantifier} of entanglement chosen here is unable to do so (in analogy with what is found running tests of genuinely multipartite non-locality\cite{CampbellPRA10}). A more refined statistical analysis is necessary for this task, as shown by the success achieved in using the variance of the entanglement distribution.

\section{Conclusions and Further Developments}
\label{conc}
We have studied the behavior of general quantum correlations shared by the elements of a spin system governed by a transverse Ising model under the influences of a collective magnetic field. Our approach was multifaceted: on one hand, we aimed at showing that some degree of care is required in the choice of the indicator of non-classicality in such a model. Naive choices dictated by the easiness of calculation implied by basis-dependent figures of merit may well lead to misleading physical conclusions. On the other hand, we have embarked in a quantitative investigation on the content of multipartite quantum correlations across the spins of a given ring prepared in its collective ground state, finding that the deep structural changes occurring close to the ring's critical point are well revealed by a multipartite extension of quantum discord. Such conclusions have been confirmed by a statistical analysis of the way entanglement is distributed across a ring of a set size: the variance of such distribution agrees in an excellent way with the predictions of global discord, while the average degree of entanglement is basically oblivious to criticality, similarly to the behavior of non-locality indicators for the very same system\cite{CampbellPRA10}. Our analysis paves the way to a more extensive study addressing thermally-affected states and aiming at relating the behavior of global discord to the most intimate statistical properties of the spin system studied here\cite{post}.
\section*{Acknowledgments}

This work was supported by the Department of Employment and Learning, the Magnus Ehrnrooth Foundation and the UK EPSRC (EP/G004759/1).

\vspace*{-6pt}   

\end{document}